\documentclass[conference]{IEEEtran}
\IEEEoverridecommandlockouts
\usepackage{cite}
\usepackage{amsmath,amssymb,amsfonts}
\usepackage{algorithm}
\usepackage{algorithmic}
\usepackage{tikz}
\usepackage{graphicx}
\usepackage{textcomp}
\usepackage{xcolor}
\usepackage{float}
\def\BibTeX{{\rm B\kern-.05em{\sc i\kern-.025em b}\kern-.08em
    T\kern-.1667em\lower.7ex\hbox{E}\kern-.125emX}}
\begin{document}
\newcommand{\squeezeup}{\vspace{-0.05mm}}
\newcommand{\tikzcircle}[2][red,fill=red]{\tikz[baseline=-0.5ex]\draw[#1,radius=#2] (0,0) circle ;}%

\title{Parallelized Instantaneous Velocity and Heading Estimation of Objects using Single Imaging Radar\\}

\author{\IEEEauthorblockN{Nihal Singh\IEEEauthorrefmark{1},
Dibakar Sil\IEEEauthorrefmark{2}, and
Ankit Sharma\IEEEauthorrefmark{2}}
\IEEEauthorblockA{Steradian Semiconductors\\
Email: \IEEEauthorrefmark{1}nihal.s.singh@gmail.com,
\IEEEauthorrefmark{2}dibakars@steradiansemi.com,
\IEEEauthorrefmark{2}ankits@steradiansemi.com}}

\maketitle

\begin{abstract}
The development of high-resolution imaging radars introduce a plethora of useful applications, particularly in the automotive sector. With increasing attention on active transport safety and autonomous driving, these imaging radars are set to form the core of an autonomous engine. One of the most important tasks of such high-resolution radars is to estimate the instantaneous velocities and heading angles of the detected objects (vehicles, pedestrians, etc.). Feasible estimation methods should be fast enough in real-time scenarios, bias-free and robust against micro-Dopplers, noise and other systemic variations. This work proposes a parallel-computing scheme that achieves a real-time and accurate implementation of vector velocity determination using frequency modulated continuous wave (FMCW) radars. The proposed scheme is tested against traffic data collected using an FMCW radar at a center frequency of 78.6 GHz and a bandwidth of 4 GHz. Experiments show that the parallel algorithm presented performs much faster than its conventional counterparts without any loss in precision.

\end{abstract}


\section{Introduction}

High-resolution imaging radars are touted as the eyes of an autonomous engine by major automotive companies and organizations around the globe \cite{dense24}. While these radars are typically proposed for geoscience applications\cite{brenner2008sarradar}, traffic monitoring \cite{bovolo2012hierarchicaltrafficmoni}, and military surveillance~\cite{sun2020mimoservillence}, increasing attention on active transport safety has shifted the focus towards advanced driver assistance systems (ADAS) and autonomous driving.

\begin{figure*}[htb]
    \centering
    \centerline{\includegraphics[width = 488pt]{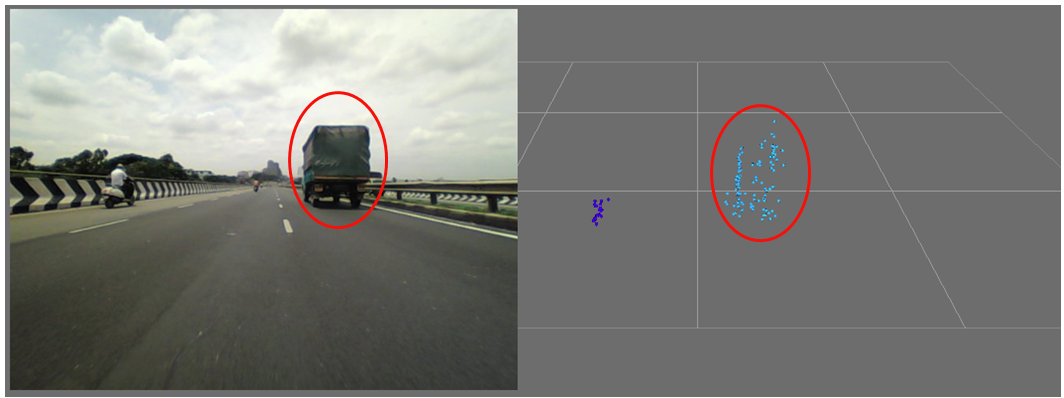}}
    \caption{[Left] Camera Image depicting a subset of the captured Radar Frame's Field of View, [Right] Visualization of generated 3D Point Cloud (78.6 GHz center freq., 4GHz BW)}
    \label{fig:camerapc}
\end{figure*}

Frequency-Modulated Continuous Wave (FMCW) imaging radars carry out sensing and monitoring of the environment by determining various attributes of surrounding objects. This process of object detection entails measuring differences in frequency and/or phase between the transmitted signals, and received echoes. Initial physical layer radar data processing results in a point cloud (PC) (refer Fig.~\ref{fig:camerapc}) with each point including attributes such as range, velocity, azimuth angle, elevation angle, etc., with respect to the observing radar's axes of reference.

\begin{figure}[hbt]
    \centering
    \centerline{\includegraphics[width = 200pt]{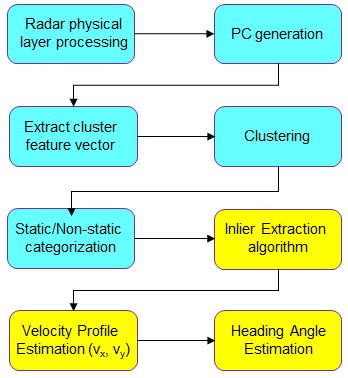}}
    \caption{Algorithm Chain for Radar Processing Stack up until Heading Angle Estimation}
    \label{fig:flowchart}
\end{figure}

\begin{figure*}[htb]
    \centering
    \centerline{\includegraphics[width = 488pt]{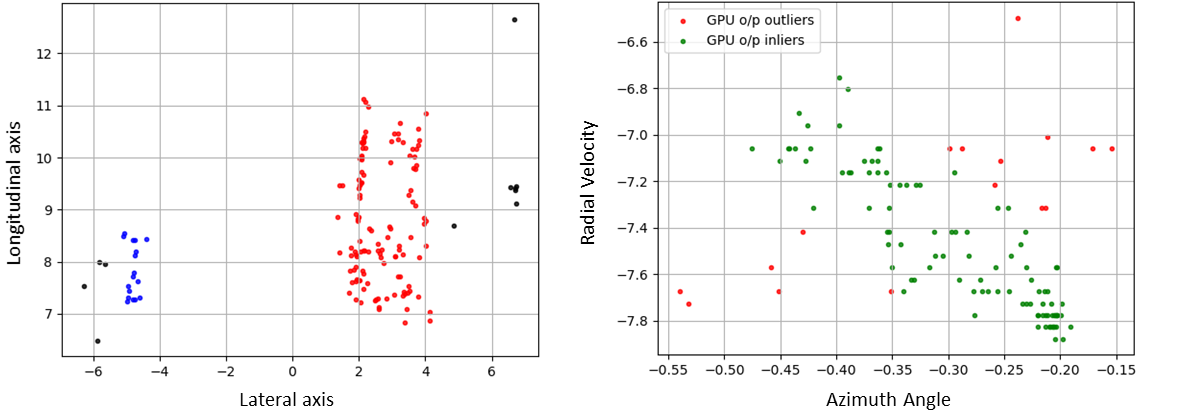}}
    \caption{[Left] 2D Clustering output along Longitudinal and Lateral axes. Cluster points corresponding to Truck is represented with (\tikzcircle{2pt}), cluster corresponding to motorcycle is respresented with (\tikzcircle[blue, fill=blue]{2pt}), and noise points are shown with (\tikzcircle[black, fill=black]{2pt}), [Right] Result of inlier detection on cluster points corresponding to the truck. (78.6 GHz center freq., 4GHz BW)}
    \label{fig:clusterransac}
\end{figure*}

Within the PC, an object appears as a dense collection of points, also termed as a cluster. As represented in Fig.~\ref{fig:flowchart}, the step after PC generation is the clustering feature vector extraction. This process comprises selection of attributes to construct a feature vector, to base the clustering process on. Typically, the feature vector includes the Cartesian counterparts of the detected ranges but could include other attributes of the detected points as well. Subsequently, clustering involves grouping of points similar to each other in the feature vector sense into respective clusters (refer Fig.~\ref{fig:clusterransac}). This is usually carried out by employing spatial clustering algorithms such as Nearest Neighbour \cite{nearestneighbour1975}, DBSCAN \cite{ester1996dbscan}, Grid-DBSCAN \cite{darong2012griddbscan} etc. After segregating the PC into clusters, it is desired to determine the velocity profile and heading of each such cluster.

Recent literature shows that separation of closely spaced clusters by standard density-based methods gives unreliable categorizations of certain points. 
Furthermore, these algorithms are not noise robust and require further developments for accurate clustering in applications utilizing real radar data~\cite{2020latestclustering}. Oscillatory motion of sub-parts of an object induces modulations in the primary reflected signal, termed as micro-Doppler. A common example of this is the signal produced by rotating wheels of vehicles in motion \cite{chen2019microDoppler}. Accurate estimation of the instantaneous velocity and heading angle is thus hindered by the influence of these effects.

Typical velocity profile estimation by radar is achieved by using more than one sensor \cite{kellner2014instantaneousfullmotion} or restricting the detection to only linear motion \cite{kellner2013instantaneous}. 
Another commonly used approach is tracking \cite{kellner2015tracking}, though tracking solutions take multiple iterations of range and projected velocity determination to converge to the object's correct vector velocity, and therefore are not instantaneous.
In \cite{lidman2018clustering}, Lidman \emph{et al}. have proposed an algorithm to estimate the instantaneous velocity by extracting the shape of the cluster. However, this approach is only feasible when measurements from multiple Doppler radars are combined for accurate shape formation. 

This paper proposes and evaluates instantaneous velocity profile and heading estimation of objects in real scenarios using a single Multiple-Input and Multiple-Output (MIMO) FMCW imaging radar.
The main contributions of this work are the real-time implementations of:
\begin{itemize}
    \item Inlier determination algorithm for the purpose of real-time filtering of random and systemic data variations from the radar detected Doppler velocity and azimuth angle.
    \item Least squares approximation method to simultaneously compute instantaneous vector velocities of multiple objects. 
\end{itemize}

Section~$2$ describes the problem statement and motivates the need for a better real-time solution to determine instantaneous vector velocity. 
Section~$3$ explains the proposed parallelized solution. 
Section~$4$ highlights the results obtained by employing the proposed method and Section~$5$ provides the concluding remarks.

\section{Problem Formulation}

For practical radar applications in the automotive sector, accurate and real-time determination of detected objects' attributes is essential. Work presented in this paper uses a single MIMO imaging radar module from Steradian Semiconductors~\cite{stersemi} to capture radar data in common automotive scenarios.

Range, Doppler speed and angle of arrival (AoA) from the said radar are calculated using frames of data transmitted and echoed back to the radar. Each frame is composed of a set of frequency modulated chirps, where the inter-chirp time is $T_c$, and $N_c$ such chirps constitute a single frame. Pulse compression with the received echo produces an intermediate frequency (IF) signal that gives range information within the digital baseband samples of a chirp, Doppler information across chirps' samples and AoA information from samples across antenna elements~\cite{pulsecomp}.

With the starting frequency of a chirp signal as $f_0$, and slope $S_u$, instantaneous frequency of the chirp signal can be written as

\begin{equation}
    f(t) = f_0 + S_ut, \quad 0<t<T_c
\end{equation}
where $T_c$ is the chirp ON time. Phase within the support of $t$ can be represented by 
\begin{align}
    \phi(t) & = \int_0^t 2 \pi f(t)dt \nonumber \\
    & = 2 \pi (f_0t + \frac{1}{2}S_ut^2).
\end{align}
and the final transmitted chirp can be written as
\begin{align}
    s(t) & = e^{\mathrm{j}\phi(t)} \nonumber \\
    & = e^{\mathrm{j} 2 \pi(f_0t + \frac{1}{2}S_ut^2)}.
\end{align}

After the formation of clusters, the main task at hand is to estimate and assign a velocity profile and heading to real life objects represented by said clusters. Extensive testing on radar data had led to the observation that standard estimation methods are not inherently immune to the effects of micro-Dopplers, multipath reflections \cite{feger2012multichannel}, clustering inaccuracies, etc. Robust determination of object velocity and heading thus necessitates an inlier estimation algorithm as a prerequisite for the purpose of noise elimination \cite{kellner2013instantaneous, kellner2014instantaneousfullmotion}. 
Furthermore, standard inlier subset detection algorithms like RAndom SAmple Consensus (RANSAC)~\cite{derpanis2010ransacoverview} are iterative in nature and scale with both the number of points in each cluster, and the numbers of clusters itself~\cite{chum2008optimalransac, schnabel2007efficient}.

Radar data captured in urban scenarios has shown that the number of detected clusters in the PC vary significantly with the infrastructure and prevalent traffic conditions. Existing radar processing algorithms determine the velocity profile and heading angle of a single object at a time, resulting in drastic variation of total processing time based on the neighbourhood of the ego-vehicle.

Hence, an end-to-end time-optimized algorithm is needed to increase the throughput of the radar processing stack and improve crash avoidance and collision warning for advanced driver assistance systems and autonomous vehicles. This work discusses such an algorithm that leverages the power of graphics processing units (GPUs) to run thousands of iterations in parallel by exploiting simultaneous thread execution capabilities.

\begin{figure*}[htb]
    \centering
    \centerline{\includegraphics[width = 494pt]{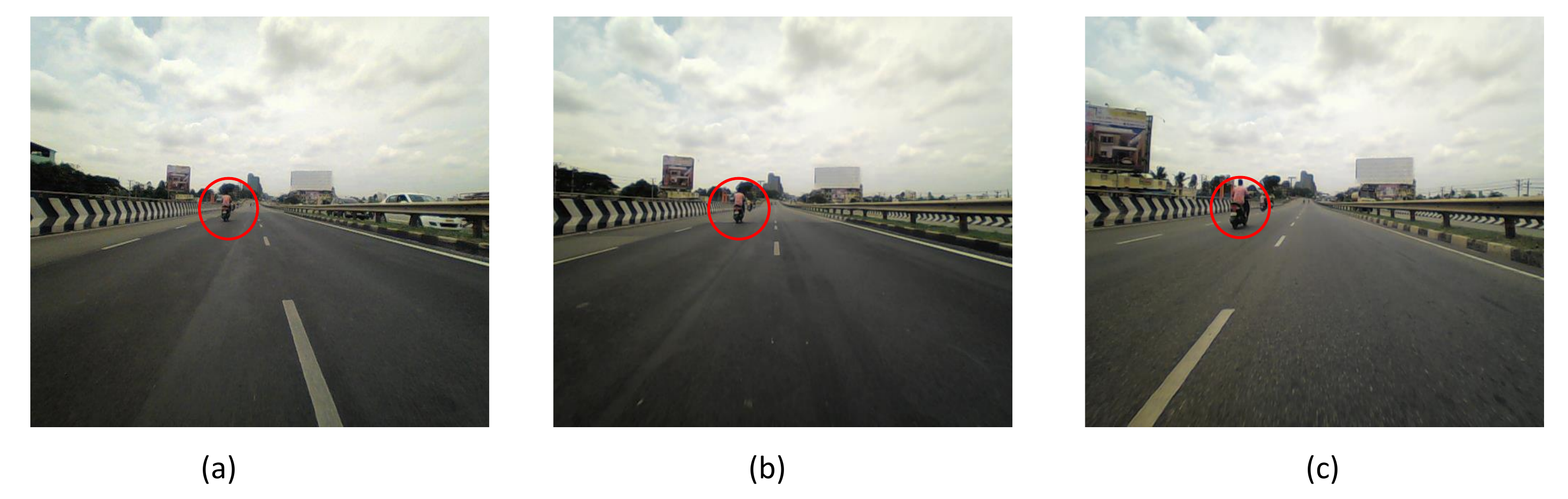}}
    \caption{Starting, Intermediate, and Final Camera Images corresponding to frames resulting in undesired values for LSQ only solution. Radar data was captured by a vehicle accelerating towards a motorbike (78.6 GHz center freq., 4GHz BW)}
    \label{fig:camerapeaks}
\end{figure*}

\begin{figure*}[htb]
    \centering
    \centerline{\includegraphics[width = 498pt]{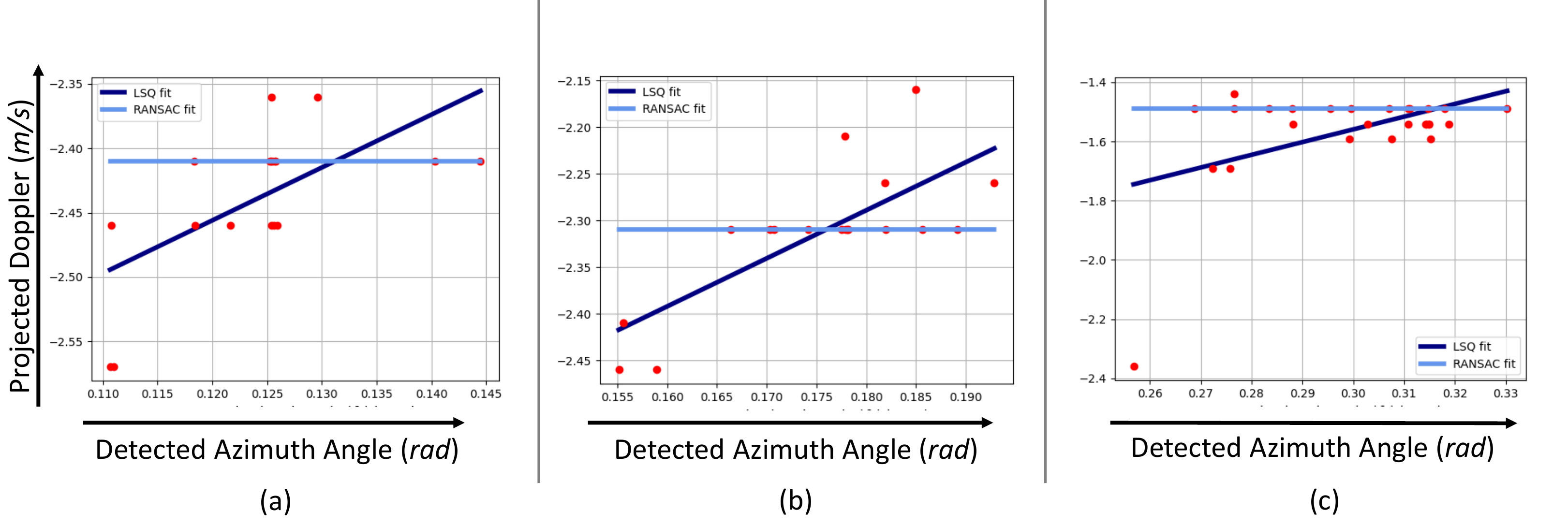}}
    \caption{Comparison between LSQ only, and RANSAC plus LSQ fits on unfiltered input data corresponding to the frames represented by the camera images above. Performing RANSAC on the input dataset with a threshold of MAD results in inlier categorization of points corresponding to the true projected Doppler of the motorbike in the camera frames of Fig~\ref{fig:camerapeaks}. Without this filtering pre-requisite, LSQ alone is skewed easily by the presence of varying projected Dopplers.  (78.6 GHz center freq., 4GHz BW)}
    \label{fig:lsqransacpeacks}
\end{figure*}


\section{Proposed Solution}
The two proposed solutions for parallelized RANSAC inlier detection and the subsequent vector velocity detection are detailed in the following subsections.
\subsection{Parallelized RANSAC Algorithm}

The rapid development of high-performance manycore GPUs has led to the widespread access of computing platforms with increasing instruction throughputs. These platforms have enabled parallel-computing based algorithms to become an attractive option for practical radar processing applications.
Fig.~\ref{fig:clusterransac} re-iterates the need for an inlier estimation algorithm by illustrating how velocity and heading determination can be heavily skewed by the presence of modulated frequency components in the input dataset. 

The algorithm discussed in this section performs simultaneous categorization of all points in the PC as either inliers or outliers of their respective clusters. This concurrency is achieved by scheduling numerous thread instances, with each thread corresponding to a linear fit generated by a pair of seed points. While $N\_clusts$ represents the number of clusters detected in a frame, the $max\_trials$ parameter determines the number of such fits evaluated for each cluster, and is usually adjusted based on the hardware specification of the user's platform (\ref{eq:threadcount})
\begin{equation}
\label{eq:threadcount}
\begin{split}
thread\_count =  max\_trials \times N\_clusts
 \end{split}
\end{equation}
where $thread\_count$ represents the number of threads that the GPU can schedule in parallel without significant increase in execution time. This reconstructed version of the standard RANSAC algorithm operates on the projected velocity ($\mathbf{v_{r}}$) and azimuth angle ($\theta$) of all points belonging to formed clusters.

\medskip
\begin{algorithm}
\caption{RANSAC GPU Kernel}
\begin{algorithmic}
\REQUIRE Random no.s spanning indices for every cluster
\ENSURE $tIdx \leq max\_trials \times N\_clusts$
\STATE $curr\_cluster \leftarrow tIdx / N\_clusts$
\STATE $seedpt1 \leftarrow $ $normalize$($data[RND1]$)
\STATE $seedpt2 \leftarrow $ $normalize$($data[RND2]$)
\STATE $m \leftarrow $ $Slope$($seedpt1,seedpt2$)
\STATE $c \leftarrow $ $Y\,intercept$($seedpt1,seedpt2$)
\FORALL {$index$ in $curr\_cluster$}
\STATE $test\_pt \leftarrow $ norm ($data[index]$)
\STATE $aa = -m$, $bb = 1$, $cc = -c$
\STATE $distn = aa \times test\_pt[0]+bb \times test\_pt[1]+cc$
\STATE $distd = \sqrt{aa^2 + bb^2}$
\STATE $dist =$ abs$(distn / distd)$
\IF{$dist \leq thresh[curr\_cluster]$}
\STATE $inliers[tIdx, index] = 1$
\STATE $num[tIdx] = num[tIdx] + 1$
\ENDIF
\ENDFOR
\end{algorithmic}
\end{algorithm}

The median absolute deviation (MAD) of normalized $\mathbf{v_{r}}$ values determines the thresholding corridor around the linear fit (\ref{eq:MAD}). This statistical parameter was selected after comparison with other known data variability measures for both real and simulated datasets. Shown below is the evaluation of this parameter, where $\mathbf{v_{r_i}}$ is the projected velocity of each cluster point, and $med(x)$, denotes the median of the $x$ vector.

\begin{equation}
\label{eq:MAD}
\begin{split}
MAD = \frac{1}{n}\sum_{i=1}^{n} [\mathbf{v_{r_i}}-med(\mathbf{v_{r}})]
\end{split}
\end{equation}

Within each thread, while two seed points are used to generate the fit hypothesis, the remaining cluster points are iterated through to calculate respective perpendicular distances to the fit.

The inlier mask for a given cluster is chosen based on the thread index corresponding to iteration that results in the largest number of post-thresholding points.
The combined inlier subset of the PC is determined based on the logical OR operation on the parallely generated inlier masks for each cluster. The result of the RANSAC filtering is much easily observed in the straight line motion of a motorbike traveling ahead of the ego-vehicle in Fig~\ref{fig:lsqransacpeacks}.

\begin{figure*}[htb]
    \centering
    \centerline{\includegraphics[width = 488pt]{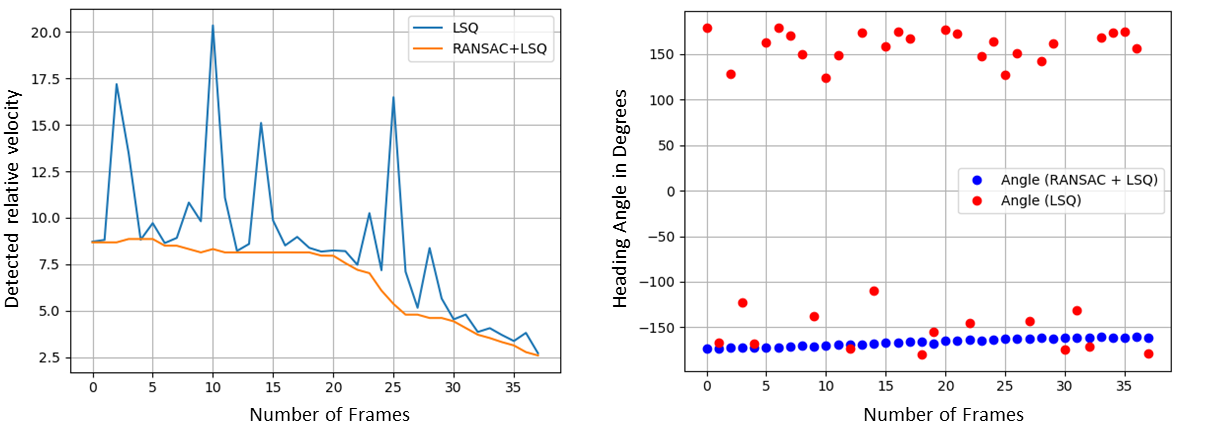}}
    \caption{[Left] Detected relative velocity plotted across frames shown in Fig~\ref{fig:camerapeaks}. Comparison is drawn between the approach implemented RANSAC plus LSQ, and LSQ only, [Right] Estimated Heading Angle is plotted across the same range of frames. Similar comparison is plotted to observe the difference in variability in the estimated angle (78.6 GHz center freq., 4GHz BW)}
    \label{fig:velandhead}
\end{figure*}

\subsection{Instantaneous velocity and Heading angle estimation by Least-Squares Kernel}
\label{ssec:subhead}

One of the important problems for ADAS is to estimate the velocity profile and heading angle of the objects detected by the ego-vehicle.
Keeping in mind the need for real-time sensing and detection, a GPU kernel is proposed to implement the least squares approximation method. This kernel instantiates a thread in parallel for each cluster to ensure optimal processing time, independent of the number of clusters.
Let, $v_x$ and $v_y$ be the velocities in the longitudinal and lateral directions, respectively (refer Fig.~\ref{fig:camerapc} for axes representation). 
Also, let $\mathbf{v_{r_i}}$ and $\mathbf{\theta_{i}}$ denote the vector of detected radial Doppler speeds and the azimuth angles of all the points on the $i^{th}$ cluster.
For any detected point $j$ on cluster $i$, we have

\begin{equation}
    {v_{r_{i,j}}} = v_{x_{i,j}}\text{cos}({\theta_{i,j}}) + v_{y_{i,j}}\text{sin}({\theta_{i,j}}),
\label{LSQ_eq}
\end{equation}

where $ j \in [1, N_i]$ with $N_i$ denoting the number of points making the $i^{th}$ cluster. Also, $v_{x_{i,j}}$ and $v_{y_{i,j}}$ denote the vector velocity components of a detected point on the cluster.
Under the rigid body assumption, each reflecting point on a cluster will have the same vector velocity. Using vector notations, the velocity vector $\mathbf{v_{i}} = (v_{x_i}, v_{y_i})^T$ is fixed for all the points on cluster $i$. Therefore, Eq.~\eqref{LSQ_eq} can be rewritten as follows:

\begin{equation}
    \mathbf{v_{r_i}} = A_i\mathbf{v_i}, \implies
    \mathbf{v_i} = A_i^\dagger \mathbf{v_{r_i}},
    \label{eq:LSQPerCluster}
\end{equation}
where

\begin{equation*}
    A_i = 
\begin{bmatrix}
\text{cos}(\theta_{i1}) & \text{sin}(\theta_{i1})\\
\text{cos}(\theta_{i2}) & \text{sin}(\theta_{i2})\\
\vdots & \vdots\\
\text{cos}(\theta_{iN_i}) & \text{sin}(\theta_{iN_i})
\end{bmatrix}.
\label{Amat}
\end{equation*}

\noindent and $A_i^\dagger$ denotes the matrix inverse of $A_i$~\cite{GenrlInv}.
Following this, we compute the heading angle $(\phi_i)$ which is defined as the arc tangent of the ratio between $v_{y_i}$ and $v_{x_i}$ with respect to the ADAS X-axis~\cite{adascoord}.
\begin{equation}
    \phi_i = \arctan (\frac{v_{y_i}}{v_{x_i}}).
\end{equation}

\begin{algorithm}
\caption{GPU kernel for component velocity $(v_x, v_y)$ and heading angle $(\phi)$ estimation}
\begin{algorithmic}
\REQUIRE $A_i$ and $v_{r_i}$ for every cluster (Eq.~\eqref{LSQ_eq}, Eq.~\eqref{eq:LSQPerCluster})
\ENSURE $i \leq num\_of\_clusters$
\STATE $\mathbf{v_i} = A_i^\dagger \mathbf{v_{ri}}$
\STATE \textit{Set $\mathbf{v_i[1]}$ as the lateral velocity}, $v_{y_i} \leftarrow \mathbf{v_i[1]}$
\STATE \textit{Set $\mathbf{v_i[0]}$ as the longitudinal velocity}, $v_{x_i} \leftarrow \mathbf{v_i[0]}$
\STATE $\phi_i \leftarrow \arctan(\frac{v_{y_i}}{v_{x_i}})$

\end{algorithmic}
\end{algorithm}

For each cluster $i$, coordinate pair $(v_{xi}, v_{yi})$ denotes the vector velocity of the cluster. Each of the cluster's vector velocity gets computed in parallel, subsequent to inlier detection using RANSAC.

\section{Results}
\label{sec:results}
The implementation proposed in this work is compared against a standard LSQ approach based on tests run on captured radar data. Both approaches were evaluated on the basis of their ability to $(1)$ operate in real-time and $(2)$ accurately determine instantaneous velocity and heading angle of detected objects.
As seen in Fig.~\ref{fig:velandhead}, the presented approach has much better overall performance with lesser erroneous spikes in the determined velocity, and no output inconsistency in estimated heading angle. The depicted overall improvement is a result of the real-time outlier eradication facilitated by the proposed parallelized RANSAC algorithm and heading estimation.

\begin{table}
\caption{RANSAC Execution Time Comparisons}
\label{tab:tresultsransac}
\begin{center}
	\begin{tabular}{||c|c|c|c||}
		\hline
		Number of & Points per & Parallelized & sklearn\\
		Clusters & Cluster & RANSAC & RANSAC\\
		\hline
		\hline
		8 & $100$ & $1.64ms$ & $70.50ms$ \\
		\hline
		8 & $150$ & $1.77ms$ & $76.58ms$ \\
		\hline
		16 & $100$ & $1.68ms$ & $141.15ms$ \\
		\hline
		16 & $150$ & $1.82ms$ & $152.97ms$ \\
		\hline
		32 & $100$ & $1.73ms$ & $281.95ms$\\
		\hline
		32 & $150$ & $1.89ms$ & $305.85ms$\\
		\hline
		64 & $100$ & $1.98ms$ & $564.48ms$\\
		\hline
		64 & $150$ & $2.23ms$ & $611.91ms$\\
		\hline
		\hline
	\end{tabular}
\end{center}
\end{table}

This proposed approach was further compared against the commonly used scikit-learn RANSACRegressor \cite{scikit-learn} for varying input conditions as shown in Table~\ref{tab:tresultsransac}. An NVIDIA GeForce GTX 1060 3GB \cite{nvidiacite} was used to run the GPU kernels discussed in Sections~$3.1$ and~$3.2$. The approach presented in this work was observed to be much faster due to its ability to perform inlier [outlier] detection on points using a twofold parallelization scheme, $(1)$ across multiple detected points on a cluster and $(2)$ across multiple clusters themselves. Scikit-learn RANSACRegressor could filter out noise from only one cluster at a time, resulting in total time taken scaling linearly with the number of clusters. With real-time applications as autonomous driving and when the number of detected clusters scale up depending on traffic scenarios, it is evident that such a twofold parallelization scheme is an extremely important feature. Similar comparison is also shown between the proposed LSQ Kernel and a standard LSQ approach using numpy.linalg.pinv \cite{harris2020array} as shown in Table~\ref{tab:tresultslsq}. This comparison also indicates the significant overall speed-up achieved by using the parallelized pseudoinverse algorithm.

\begin{table}
\caption{LSQ Execution Time Comparisons}
\label{tab:tresultslsq}
\begin{center}
	\begin{tabular}{||c|c|c|c||}
		\hline
		Number of & Points per & Parallelized & NumPy\\
		Clusters & Cluster & pseudoinverse & pseudoinverse\\
		\hline
		\hline
		8 & $100$ & $1.81ms$ & $2.08ms$ \\
		\hline
		8 & $150$ & $1.83ms$ & $2.10ms$ \\
		\hline
		16 & $100$ & $1.81ms$ & $4.18ms$ \\
		\hline
		16 & $150$ & $1.83ms$ & $4.21ms$ \\
		\hline
		32 & $100$ & $1.81ms$ & $8.74ms$\\
		\hline
		32 & $150$ & $1.83ms$ & $8.77ms$\\
		\hline
		64 & $100$ & $1.81ms$ & $16.84ms$\\
		\hline
		64 & $150$ & $1.83ms$ & $17.19ms$\\
		\hline
		\hline
	\end{tabular}
\end{center}
\end{table}

\section{Conclusion}
\label{sec:conclusion}

The algorithm presented in this paper showcases a significant speed-up of existing estimation algorithms along with supporting quantitative metrics, after extensive verification across various practical scenarios. This work demonstrates the need for robust and optimized inlier detection methods for time critical applications such as ADAS, traffic monitoring, etc. The proposed algorithm is a significant step forward in the development of practically realizable ADAS, and will scale directly with enhancements in GPU hardware and programming interfaces as well.

\end{document}